# A novel approach to the quantum Buneman instability

Herman R. Vallejos
Department of Physics, University of La Serena, La Serena, Chile.

Abstract
Consider a non collisional quantum plasma, where electrons stream with a velocity $u_0$ relative to the ions, which are assumed to be initially at rest due to their mass. A quantum dispersion relation is obtained having considered the Bohm interpretation of Quantum Mechanics by means of Quantum Magnetohydrodynamics. This quantum dispersion relation contains the classical non dimensional parameter $\kappa = k u_0 / \omega_{pe}$ and the quantum non dimensional parameter $H = \hbar \omega_{pe} / m_e u_0^2$. A quartic equation ( $Aw^4 + Bw^3 + Cw^2 + Dw + E = 0$, where w is the normalized frequency ) is obtained from the quantum dispersion relation. So the novel approach to the quantum Buneman instability consists in having solved for the first time this quartic equation in order to obtain the instability intervals.

**My next paper will show how the values of $\kappa$ and H can be replaced by the initial velocity $u_0$ in the quantum instability intervals obtained in this paper.**

## I. INTRODUCTION

This instability arises when electrons stream at a velocity $u_0$ relative to ions which are assumed to be initially at rest. There is another version of this instability, i.e. the Two stream instability which originates when there are two oppositely directed electron streams with velocities $\pm u_0$. Both versions have been analyzed classically and from the quantum point of view. A biquadratic equation ( $Aw^4 + Bw^2 + C = 0$, where w corresponds to the normalized frequency ) is obtained from the quantum dispersion equation when the quantum Buneman instability is analyzed at low frequencies and when considering two opposite electron streams ( Two stream instability ) [1].

**In this research the quantum Buneman instability is analyzed without considering low frequencies, therefore the equation obtained from the quantum dispersion relation is a quartic equation ( $Aw^4 + Bw^3 + Cw^2 + Dw + E = 0$, where w corresponds to the normalized frequency ), which is very complicated to solve algebraically, so it was solved numerically and this constitutes the novel approach to the quantum version of the Buneman instability because this equation had not been solved before.**

From Statistical Mechanics it is well known that classical systems are characterized by high temperatures and a low particle density, whereas quantum systems are characterized by low temperatures and a high particle density. But a high temperature does not always mean we have a classical system, indeed the particle density of the system being analized is the one which sets the limit bewteen a classical and a quantum system. A white dwarf is the best example, a white dwarf is at $1 \times 10^7$ ( K ) and its particle density is $1 \times 10^{36}$ ( m$^{-3}$ ), if we compute $k_B T$ we find out that $k_B T < E_{fermi}$ [2]. This result tells us that a white dwarf is a quantum plasma no matter its high temperature. Another way to demonstrate that a white dwarf is a quantum plasma is by means of the De Broglie thermal wave length $\lambda_{DB} = h/\sqrt{2\pi m k_B T}$. Quantum effects must be taken into account whenever the De Broglie thermal wave length is equal or greater than the average distance between particles. The distance between particles in plasmas is given by $d = n^{-1/3}$, where $n$ is the particle density, so we obtain $\lambda_{DB} > n^{-1/3}$ for the electrons and $\lambda_{DB} < n^{-1/3}$ for the ions. This result shows that in a white dwarf the electrons are the quantum constituents meanwhile ions are considered classical particles. Thus a white dwarf is a fermion plasma.

On the other hand, the quantum coupling parameter is given by $g_{quantum} = E_{interaction}/E_{Fermi} = 2e^2 m_e / \left[ (3\pi^2)^{2/3} \hbar^2 \epsilon_0 n^{1/3} \right]$ [3], then we can see that very dense plasmas ( *plasmas with a high particle density* ) are non collisional. That is equivalent to neglecting the energy of interaction, therefore a non collisional quantum plasma is a quantum ideal gas.

## II. QUANTUM MAGNETOHYDRODYNAMICS

Plasmas are viewed as a conducting fluid in magnetohydrodynamics. So the equations of motion of electrons and ions, along with the continuity and Maxwell equations constitute magnetohydrodynamics. Quantum magnetohydrodynamics is based on the Bohm formulation of Quantum Mechanics. In this formulation particle trajectories are always known; in other words it is a deterministic Quantum Mechanics with respect to trajectories. Thus the Bohm Potential is added to motion equations. This is the way the Bohm Potential is obtained. Let's consider the Hamilton-Jacobi equation,

$$-\frac{\partial S}{\partial t} = H \quad . \tag{1}$$

From equation (1) we may express the wave function as :

$$\Psi = \sqrt{n} \ e^{i S/\hbar} \quad . \tag{2}$$

On the other hand, we have the Schroedinger equation for a particle under the influence of an electrostatic potential $\Phi$ .

$$i\hbar\frac{\partial \Psi}{\partial t} = -\frac{\hbar^2}{2m} \nabla^2 \Psi - e\Phi\Psi \quad . \tag{3}$$

By inserting equation (2) into equation (3), we get the continuity equation from the imaginary part :

$$\frac{\partial n}{\partial t} + \nabla \cdot (n\boldsymbol{u}) = 0 \quad , \tag{4}$$

and from the real part we get the Bohm potential :

$$E_B = -\frac{\hbar^2}{2m}\frac{1}{\sqrt{n}} \nabla^2 \sqrt{n} \tag{5}$$

It may be seen that the continuity equation remains unchanged through the quantum regime. The Bohm Potential is added to the motion equation of electrons. So the equation of motion for the electrons is :

$$n_e m_e \left(\frac{\partial \boldsymbol{u}_e}{\partial t} + \boldsymbol{u}_e \cdot \nabla \boldsymbol{u}_e\right) = e n_e \nabla \Phi + n_e \frac{\hbar^2}{2 m_e} \nabla \cdot \left(\frac{1}{\sqrt{n_e}} \nabla^2 \sqrt{n_e}\right) \quad , \tag{6}$$

and the equation of motion for the ions is :

$$n_i m_i \left(\frac{\partial \boldsymbol{u}_i}{\partial t} + \boldsymbol{u}_i \cdot \nabla \boldsymbol{u}_i\right) = -e n_i \nabla \Phi \quad . \tag{7}$$

As mentioned earlier, ions are so heavy that their De Broglie thermal wave length is less than the average distance between ions, that is why the *ion* equation of motion obeys Classical Mechanics. As part of the Maxwell equations we have the Poisson equation

$$\epsilon_0 \nabla \cdot \boldsymbol{E} = e(n_i - n_e) \quad . \tag{8}$$

**III. QUANTUM DISPERSION RELATION**

Considering a cold ( $k_B T_i \approx 0$, $k_B T_e \approx 0$ ), non collisional quantum plasma in which $B_0 = 0$, where $B_0$ is the ambient magnetic field, or if $B_0 \neq 0$ then particles are assumed to stream along the magnetic field so that their motion is not affected by the magnetic field [4], the electrons stream with a velocity $u_0$ relative to the ions, which are assumed to be initially at rest due to their mass. Under these considerations the quantum dispersion relation is obtained as follows :

III$_1$) It is assumed that oscillations are in the form of plane waves, and we will consider that they propagate along the X axis, so the electric field $E$ and $k$ lie along the X axis because we must remember that $B_0 = 0$, and so we have electrostatic oscillations, so that is why $E \parallel k$ . By having assumed propagations in the form of plane harmonic waves, we have :

$$\frac{\partial}{\partial t} \to -i\omega \quad , \quad \frac{\partial}{\partial x} \to ik \quad , \quad \frac{\partial^3}{\partial x^3} \to -ik^3 \quad . \tag{9}$$

III$_2$) Perturbation analysis : The subindices 0 and 1 correspond to equilibrium and perturbed quantities, respectively, so that we have :

$n = n_0 + n_1$   for particle density of electrons and ions
$\boldsymbol{E} = \boldsymbol{E}_0 + \boldsymbol{E}_1$   for the electric field, where $\boldsymbol{E}_0 = \boldsymbol{0}$
$\boldsymbol{u}_e = \boldsymbol{u}_{0e} + \boldsymbol{u}_{1e}$   for electron velocity
$\boldsymbol{u}_i = \boldsymbol{u}_{0i} + \boldsymbol{u}_{1i}$   for ion velocity, where $\boldsymbol{u}_{0i} = \boldsymbol{0}$   because ions are assumed to be initially at rest.

Taking into account III$_1$) and therefore that $E\hat{\boldsymbol{x}} = -\nabla \Phi$ , so from equation(6) we have that the equation of motion for the electrons becomes :

$$n_e m_e \left( \frac{\partial u_e}{\partial t} + u_e \frac{\partial u_e}{\partial x} \right) = -e n_e E + n_e \frac{\hbar^2}{2 m_e} \frac{\partial}{\partial x} \left( \frac{1}{\sqrt{n_e}} \frac{\partial^2 \sqrt{n_e}}{\partial x^2} \right) \quad , \tag{10}$$

and from equation (7) the equation of motion for the ions becomes :

$$n_i m_i \left( \frac{\partial u_i}{\partial t} + u_i \frac{\partial u_i}{\partial x} \right) = e n_i E \quad . \tag{11}$$

And from equation (4) we get the continuity equation for the electrons

$$\frac{\partial n_e}{\partial t} + \frac{\partial}{\partial x}(n_e u_e) = 0 \quad , \tag{12}$$

and again from equation (4) we get the continuity equation for the ions

$$\frac{\partial n_i}{\partial t} + \frac{\partial}{\partial x}(n_i u_i) = 0 \quad . \tag{13}$$

By means of III$_1$) and III$_2$ ) equation (8) becomes :

$$\epsilon_0 \frac{\partial E_1}{\partial x} = e(n_{1i} - n_{1e}) \quad . \tag{14}$$

In other words, equation (14) corresponds to equation (8) being considered to first order. This is what is called the linearization of equation (8). Using equations (9) along with linearizing equations (10) and (12) we get $n_{1e}$ ( where $\mathbf{u_{1e}} \cdot \nabla \mathbf{u_{0e}} = 0$ in equation (6) because $\mathbf{u_{0e}}$ is assumed to be uniform [4] ) and by doing the same procedure with equations (9), (11) and (13) we get $n_{1i}$ and so we finally get the quantum dispersion relation from equation (14) :

$$1 = \frac{\omega_{pi}^2}{\omega^2} + \frac{\omega_{pe}^2}{\left(\omega - k\, u_{oe}\right)^2 - \frac{\hbar^2 k^4}{4 m_e^2}} \quad , \tag{15}$$

where $\omega_{p\alpha} = \sqrt{n_0 e^2 / m_\alpha \varepsilon_0}$ is the plasma frequency of $\alpha = electrons, ions$

Equation (15) may be written as :

$$1 = \frac{1}{\left(\frac{\omega^2}{\omega_{pi}^2}\right)} + \frac{1}{\left(\frac{\omega}{\omega_{pe}} - \frac{k\, u_{oe}}{\omega_{pe}}\right)^2 - \frac{\hbar^2 k^4}{4 m_e^2 \omega_{pe}^2}} \quad . \tag{16}$$

Considering $\frac{m_e}{m_i} \approx 5 x 10^{-4}$ we have :

$$\omega_{pi}^2 \approx 5 x 10^{-4} \omega_{pe}^2 \quad . \tag{17}$$

Inserting equation (17) into equation (16), we get :

$$1 = \frac{1}{\frac{\omega^2}{5 x 10^{-4} \omega_{pe}^2}} + \frac{1}{\left(\frac{\omega}{\omega_{pe}} - \frac{k\, u_{oe}}{\omega_{pe}}\right)^2 - \frac{\hbar^2 k^4}{4 m_e^2 \omega_{pe}^2}} \quad . \tag{18}$$

And if we set $w = \omega / \omega_{pe}$ , this is what is called the normalization of the frequency. So equation (18) becomes :

$$1 = \frac{1}{\frac{w^2}{5 x 10^{-4}}} + \frac{1}{\left(w - \frac{k\, u_{oe}}{\omega_{pe}}\right)^2 - \frac{\hbar^2 k^4}{4 m_e^2 \omega_{pe}^2}} \quad . \tag{19}$$

Considering the classical parameter

$$\kappa = \frac{k u_{0e}}{\omega_{pe}} \quad , \tag{20}$$

and the quantum parameter

$$H = \frac{\hbar \omega_{pe}}{m_e u_{0e}^2} \quad . \tag{21}$$

So equation (19) becomes

$$1 = \frac{1}{2 \times 10^3 w^2} + \frac{1}{(w-\kappa)^2 - \frac{H^2 \kappa^4}{4}} \quad . \tag{22}$$

Equation (22) is the normalized quantum dispersion relation. From the normalized quantum dispersion relation we get the classical dispersion relation by doing H=0.

$$1 = \frac{1}{2 \times 10^3 w^2} + \frac{1}{(w-\kappa)^2} \quad . \tag{23}$$

So equation (23) represents the classical dispersion relation for the Buneman instability.

We know that instabilities are related to complex frequencies. Equation (23) has four roots, so instabilities will consist of two real solutions and two complex solutions ( one conjugate of one another). To appreciate this graphically, equation (23) will be written as :

$$F(w) = \frac{1}{2 \times 10^3 w^2} + \frac{1}{(w-\kappa)^2} \quad . \tag{24}$$

In general when graphing equation (24) we get a local minimum. Let's call $w_0$ the domain at which F(w) has this local minimum. When $F(w_0) > 1$ we get two real roots and two complex roots, in other words, for $F(w_0) > 1$ we get an instability. When $F(w_0) < 1$ we get four real roots, in other words, in this case we get no instability. The next graphs will show both situations :

Let's graph equation (24) when $\kappa = 0.5$

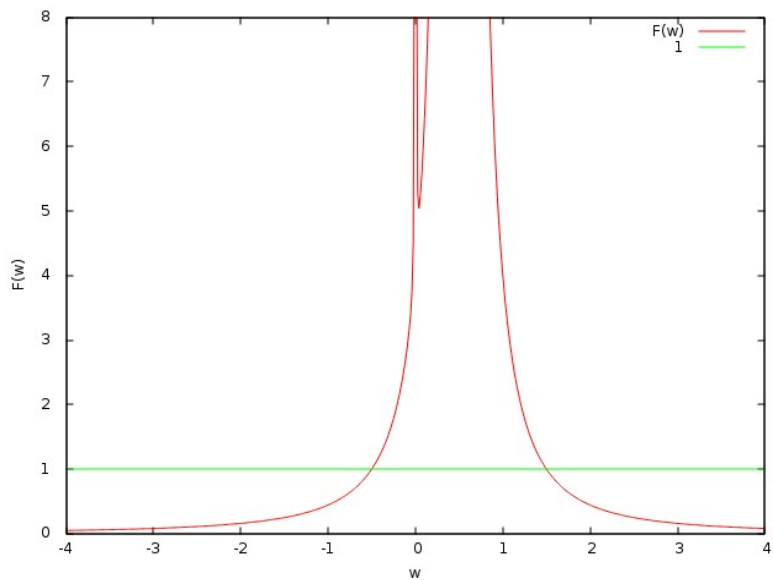

**Figure 1** F(w) is intersected twice. So we get two real roots and two complex roots, which means an instability. This is due to the fact that $F(w_0) > 1$ .

Let's graph now equation (24) when $\kappa = 1.5$

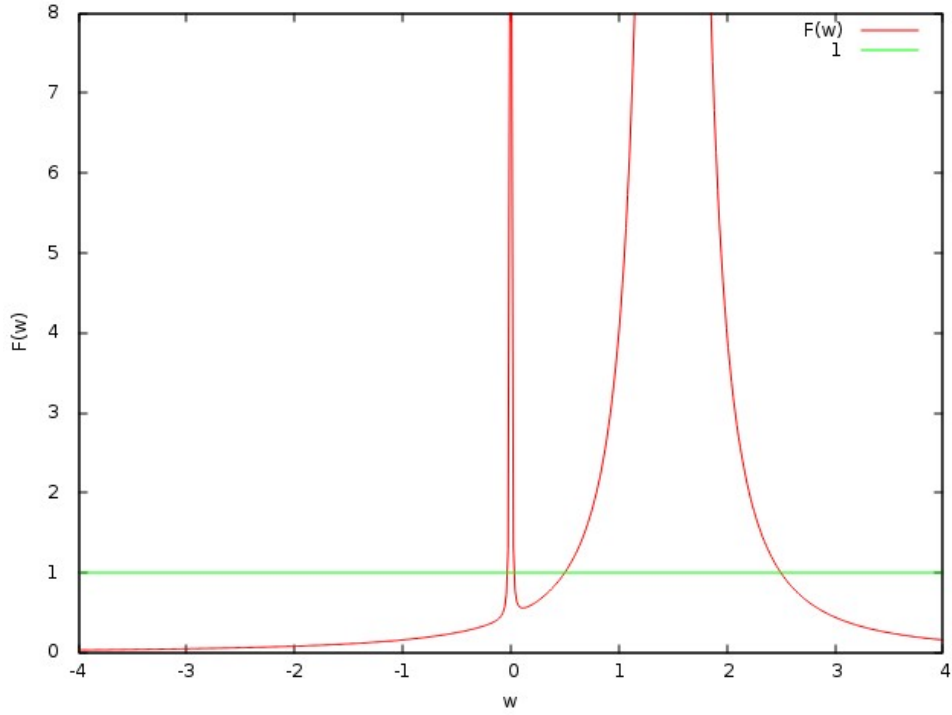

**Figure 2** F(w) is intersected four times. So we get four real roots, which means there is no instability in this case. This is due to the fact that $F(w_0) < 1$ .

Let's calculate the local minimum of F(w) in equation (24), so we get :

$$w_0 = 0.073534 \, \kappa \qquad . \tag{25}$$

Inserting equation (25) into equation (24) leads us to :

$$F(w_0) = \frac{1.25751}{\kappa^2} \tag{26}$$

From Figure 1 we see that we get instabilities whenever $F(w_0) > 1$ . Imposing this condition on equation (26) we have :

$$0 < \kappa < 1.12139 \qquad . \tag{27}$$

Equation (27) represents the classical instability intervals for the Buneman instability.

Classically in the Buneman instability, instabilities appear when $\kappa$ obeys the equation [5] :

$$\kappa < \left[1 + \left(\frac{m_e}{m_i}\right)^{1/3}\right]^{3/2} \qquad . \tag{28}$$

If we take $m_i = m_{proton}$ , we get :

$$\frac{m_e}{m_i} \approx 0.0005 \quad . \quad (29)$$

By inserting equation (29) into equation (28) we get equation (27). This shows that equation (27) is totally correct. Therefore from equation (22) we can get the classical instability intervals for the Buneman instability by doing $H \to 0$ .

## IV. QUANTUM INSTABILITY INTERVALS OBTAINED NUMERICALLY

From equation (22) ( *the normalized quantum dispersion relation* ) we get :

$$w^4 - 2\kappa w^3 - 0.25\left[H^2\kappa^4 - 4(\kappa^2 - 1)\right]w^2 + 0.001\kappa w + 1.25 \times 10^{-4}\left[H^2\kappa^2 - 4\right]\kappa^2 = 0 \quad . \quad (30)$$

This is the above-mentioned quartic equation obtained from the normalized quantum dispersion relation. This equation had never been solved before, and in the next table I show some of the numerical solutions, which represent the instability intervals.

| H | Instability Intervals |
|---|---|
| H=0.10 | $0<\kappa\leqslant 1.12$ , $19.93\leqslant\kappa\leqslant 19.99<2/H$ |
| H=0.11 | $0<\kappa\leqslant 1.12$ , $18.11\leqslant\kappa\leqslant 18.17<2/H$ |
| H=0.12 | $0<\kappa\leqslant 1.12$ , $16.59\leqslant\kappa\leqslant 16.66<2/H$ |
| H=0.13 | $0<\kappa\leqslant 1.12$ , $15.30\leqslant\kappa\leqslant 15.38<2/H$ |
| H=0.14 | $0<\kappa\leqslant 1.12$ , $14.20\leqslant\kappa\leqslant 14.28<2/H$ |
| H=0.15 | $0<\kappa\leqslant 1.12$ , $13.24\leqslant\kappa\leqslant 13.33<2/H$ |
| H=0.16 | $0<\kappa\leqslant 1.12$ , $12.41\leqslant\kappa\leqslant 12.49<2/H$ |
| H=0.17 | $0<\kappa\leqslant 1.12$ , $11.67\leqslant\kappa\leqslant 11.76<2/H$ |
| H=0.18 | $0<\kappa\leqslant 1.12$ , $11.01\leqslant\kappa\leqslant 11.10<2/H$ |
| H=0.19 | $0<\kappa\leqslant 1.12$ , $10.42\leqslant\kappa\leqslant 10.52<2/H$ |
| H=0.20 | $0<\kappa\leqslant 1.12$ , $9.89\leqslant\kappa\leqslant 9.99<2/H$ |
| H=0.21 | $0<\kappa\leqslant 1.13$ , $9.41\leqslant\kappa\leqslant 9.52<2/H$ |
| ---------- | ------------------------------------------ |
| H=0.82 | $0<\kappa\leqslant 1.38$ , $1.95\leqslant\kappa\leqslant 2.43<2/H$ |
| H=0.83 | $0<\kappa\leqslant 1.40$ , $1.90\leqslant\kappa\leqslant 2.40<2/H$ |
| H=0.84 | $0<\kappa\leqslant 1.42$ , $1.84\leqslant\kappa\leqslant 2.38<2/H$ |
| H=0.85 | $0<\kappa\leqslant 1.46$ , $1.78\leqslant\kappa\leqslant 2.35<2/H$ |
| H=0.86 | $0<\kappa\leqslant 1.50$ , $1.70\leqslant\kappa\leqslant 2.32<2/H$ |
| H≥0.87 | $0<\kappa<2/H$ |

**Table 1** This table shows the instability intervals for differente values of H and $\kappa$ . It can be seen that as H increases, the instability intervals get wider.

Equation (22) has four roots, so instabilities will consist of two real solutions and two complex solutions ( one conjugate of one another). To appreciate this graphically, equation (22) will be written as :

$$F(w) = \frac{1}{2 \times 10^3 w^2} + \frac{1}{(w-\kappa)^2 - \frac{H^2 \kappa^4}{4}} \quad . \tag{31}$$

Let's graph equation (31), taking into account that F(w)=1. Taking $\kappa = 1.12$ and H=0.10 from Table 1

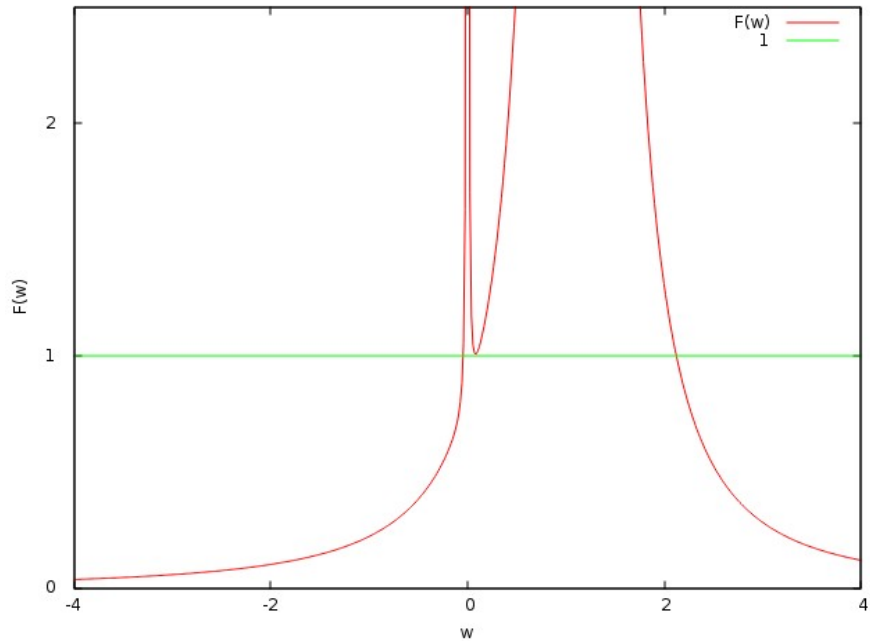

**Figure 3** F(w) is intersected just twice. This graph shows that in this case we have two real roots and two complex roots, which shows that we get an instability when H=0.10 and $\kappa = 1.12$, in total accordance with Table 1.

Graph of the quantum dispersion relation when H=0.10 and $\kappa = 1.13$ .

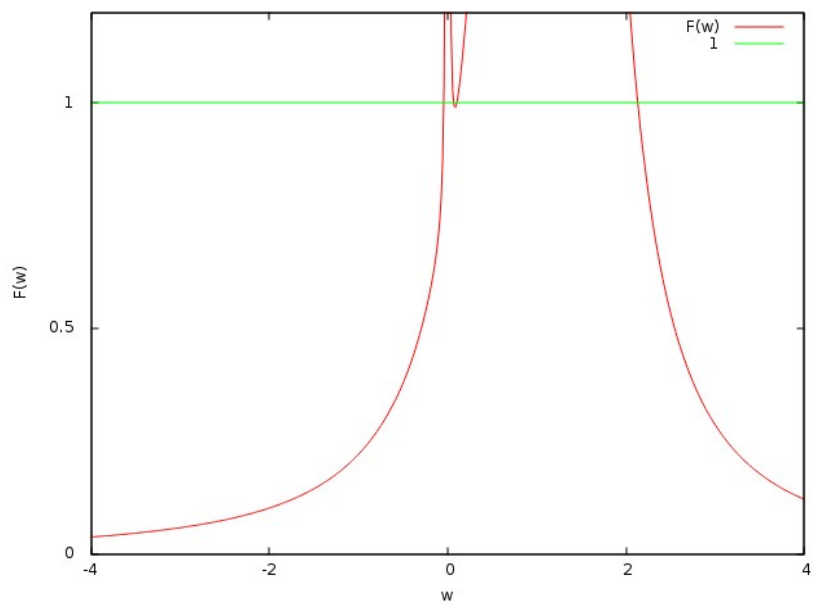

**Figure 4** F(w) is intersected four times. So we get four real roots, which means we do not have an instability when $\kappa = 1.13$ and H=0.10, which is in total accordance with Table 1.

Graph of the quantum dispersion relation when H=0.18 and $\kappa = 1$ .

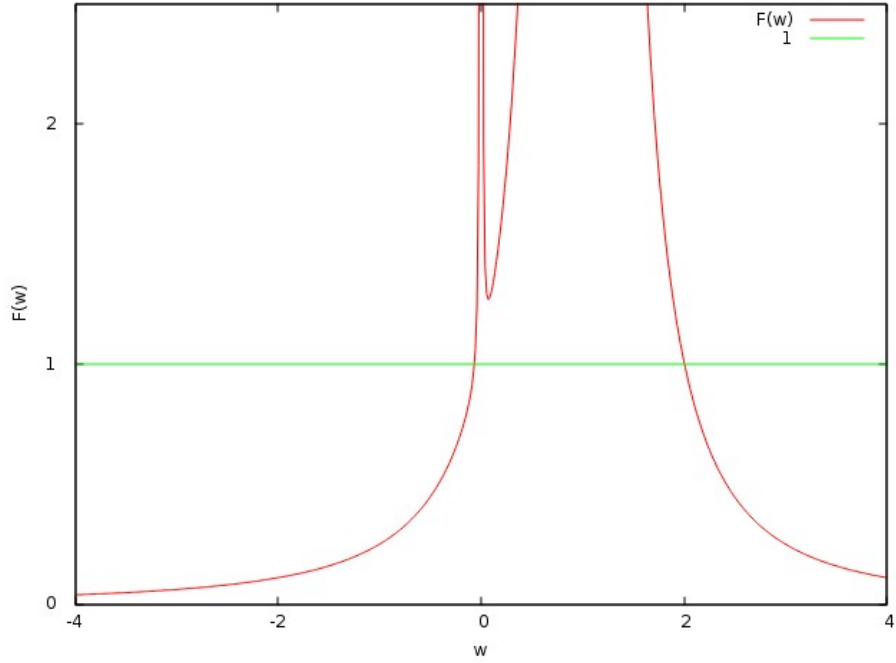

**Figure 5** F(w) is intersected twice, which means we get two real roots and two complex roots, so we have an instability when $\kappa = 1$ and H=0.18, which is in total accordance with Table 1.

Graph of the quantum dispersion relation when H=0.82 and $\kappa = 2.42$ .

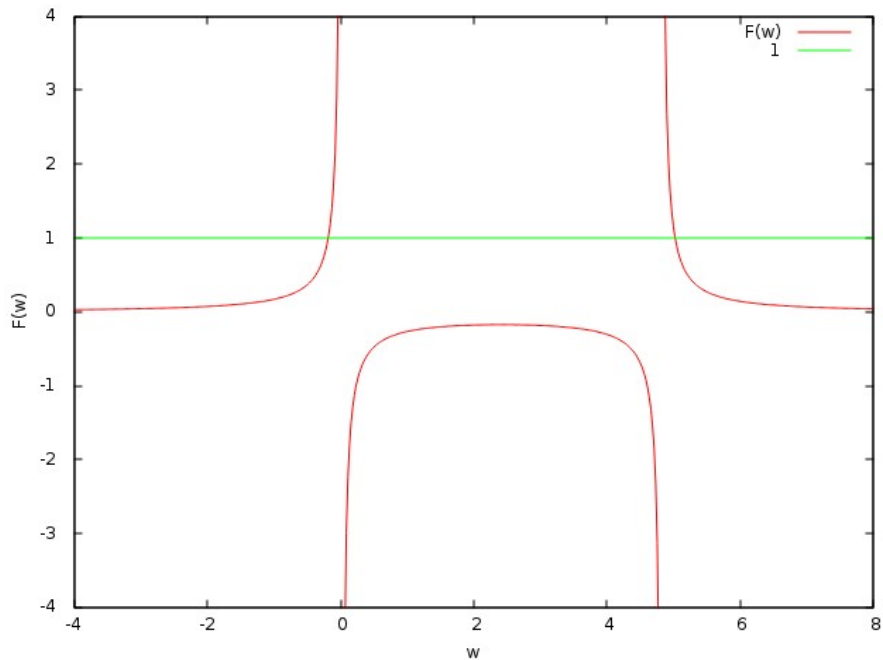

**Figure 6** F(w) is intersected twice, so we have an instability when $\kappa = 2.42$ and H=0.82, which is in total accordance with Table 1.

As already seen in section III, according to equation (27), the classical instability intervals for the Buneman appear when $\kappa \leqslant 1.12$. We will see now how the quantum parameter H modifies the classical instability intervals. To see this, let's graph equation (31) when $\kappa = 0.5$ and H = 5.

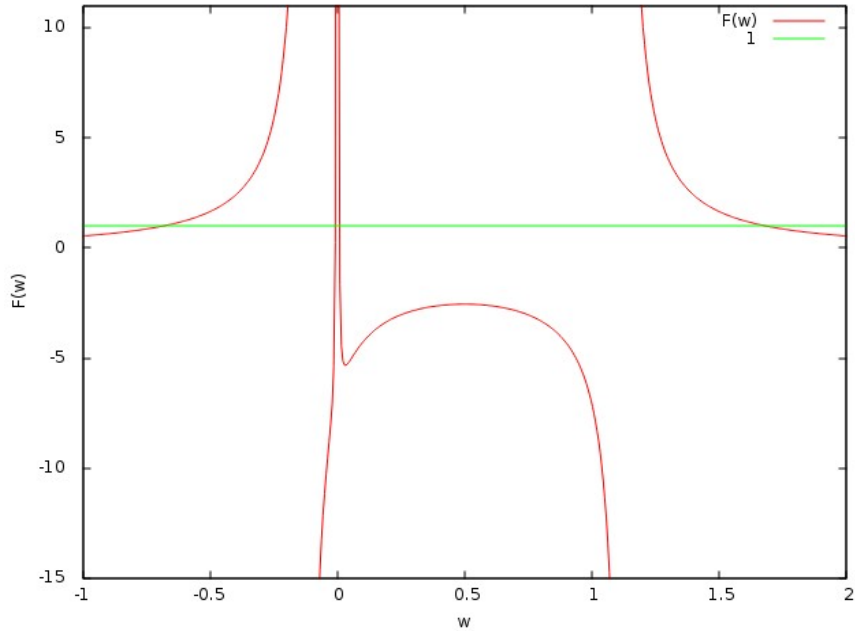

**Figure 7** F(w) is intersected four times, so in this case we get no instability even though $\kappa = 0.5$ complies with the classical condition of instabilities for the Buneman instability as ruled by equation (27).

## V. CONCLUSIONS

In the quantum Buneman instability, we have pairs of instability intervals, which does not occur classically. We know that in the classical Buneman instability we have instabilities whenever $0 < \kappa \leqslant 1.12$, and taking a look at Table 1 we see that when $H \in [0.10, 0.20]$ every pair of instability intervals contains the interval $0 < \kappa \leqslant 1.12$, so we will name these instability intervals as semi-classical instability intervals. This can be interpreted in the following way : When $H \in [0.10, 0.20]$ classical and quantum effects prevail over the instability, but quantum effects become crucial when $H \geq 0.21$. And as H increases so does the instability interval width. At the same time the instability intervals in Table 1 are wider than the instability intervals obtained in the quantum Two Stream Instability [1].

We see that the quantum parameter H modifies the classical instability intervals in the sense that if for example $\kappa = 0.5$, it does not necessarily mean that our quantum plasma will be unstable, because it all depends on the value of the quantum parameter H as we saw in section IV.